\documentclass[pre,twocolumn,showpacs]{revtex4}
\usepackage{graphicx}

\newcommand{\equ}{Eq.}

\newcommand{\fig}{Fig.}

\newcommand{\sect}{Sec.}
\newcommand{\rem}[1]{}
\newcommand{\const}{\text{constant}}
\newcommand{\dpo}{d_{\text{\footnotesize p.o.}}(E)}
\newcommand{\ddo}{d_{\text{\footnotesize d.o.}}(E)}
\newcommand{\dosc}{d_{\text{\footnotesize osc}}(E)}
\newcommand{\mod}{\,{\rm mod}\,4}
\newcommand{\FIGo}[3]{\begin{figure}[ht]%
#3%
\vspace{-0.2cm}
\caption[]{\footnotesize #2}%
\label{#1}%
\end{figure}}
	
\begin{document}

\title{Spectral properties of quantized barrier billiards}    
\author{Jan Wiersig}
\affiliation{Max-Planck-Institut f\"ur Physik komplexer Systeme, D-01187 Dresden, Germany}
\date{\today}
\email{jwiersig@mpipks-dresden.mpg.de}
\pacs{03.65.Ge, 03.65.Sq, 05.45.Mt}
\begin{abstract} 
The properties of energy levels in a family of classically pseudointegrable
systems, the barrier billiards, are investigated.
An extensive numerical study of nearest-neighbor spacing distributions, 
next-to-nearest spacing distributions, number variances, spectral form
factors, and the level dynamics is carried out. For a special member of the 
billiard family, the form factor 
is calculated analytically for small arguments in the diagonal approximation. 
All results together are consistent with the so-called semi-Poisson
statistics.
\end{abstract}
\maketitle

\section{Introduction}
\label{sec:intro}
Two decades after the first investigation of the quantum mechanics of
nonintegrable polygonal billiards~\cite{RichensBerry81}, a renewed interest in
this peculiar class of dynamical systems has shown up recently. 
One reason is the fabrication of polygonal-shaped 
optical microcavities~\cite{VKLISLA98,BGRFRKDK98,BILNSSVWW00,PCC01}. 
Another reason is the finding~\cite{BGS99,BGS01b} that certain planar
rational polygons (all angles between sides are of the form $p\pi/q$, 
where $p$, $q$ are relatively prime integers) have spectral properties 
very similar to those of mesoscopic disordered systems at the critical point
of the metal-insulator transition~\cite{SSSLS93} and to those of systems with
interacting electrons~\cite{WWP99}.  

The classical dynamics in rational polygons having at least one critical corner
with $p > 1$ is characterized as pseudointegrable~\cite{RichensBerry81}. The 
phase space is foliated by two-dimensional invariant 
surfaces~\cite{Hobson75,ZemlyakovKatok75}, like in integrable 
systems~\cite{Arnold78}, but the genus of the surfaces is larger than 
one~\cite{RichensBerry81}.    
The flow on these surfaces is typically ergodic and not
mixing~\cite{Gutkin96}. 

Pseudointegrable systems cannot be quantized according to the semiclassical
Einstein-Brillouin-Keller rule~\cite{RichensBerry81}. As a consequence, the 
statistical properties of energy levels of classically pseudointegrable 
systems are different from those of integrable systems which are generically 
well described by Poissonian random processes~\cite{BerryTabor77a}. 
For example, the nearest-neighbor spacing distribution of pseudointegrable
systems generically displays a clear level repulsion~\cite{RichensBerry81} as
in the case of the Gaussian orthogonal ensemble (GOE) of random-matrix
theory~\cite{Mehta67} which describes fully chaotic systems with time-reversal
symmetry~\cite{BGS84}. 
Significant deviations from GOE are observed first theoretically~\cite{SS93} 
and later experimentally in microwave cavities~\cite{SSSSSZ94}. 

Recently, the semi-Poisson (SP) statistics have been proposed as reference 
point for the spectral statistics of pseudointegrable 
systems~\cite{BGS99,BGS01b}. 
Following~\cite{HFS99,BGS01b}, we define SP statistics by removing every 
other level from an ordered Poisson sequence $\{x_n\}$.
Unfortunately, the term ``semi-Poisson'' was originally coined for a sequence
$\{y_n\}$ where $y_n=(x_n+x_{n+1})/2$~\cite{BGS99}. We call this here
interpolated-Poisson (IP) statistics. IP and SP statistics have identical 
nearest-neighbor spacing distributions but other spectral quantities in
general differ.

The SP conjecture has been verified numerically for right
triangular~\cite{BGS99,BGS01b} and rhombus billiards~\cite{GremaudJain98}
where only small differences to SP have been found.
However, numerical works on right triangles in a regime of high
level numbers seem to indicate that the statistical properties are 
nonstationary~\cite{Gorin01,PG01} with increasing, but still small, deviations
from SP as the energy is increased~\cite{PG01}.  
For certain right triangles, it has been shown analytically that the
spectral form factor for small arguments is located around the corresponding SP
result~\cite{BGS01}.   

However, the triangles studied in~\cite{BGS01} are not generic rational 
polygons, because they belong to the class of Veech polygons~\cite{Veech89} 
which may have special spectral properties. 
In this paper, we study the symmetric barrier
billiards~\cite{Zwanzig83,HM90,Wiersig00,Wiersig01,ZE01,EMS01,Wiersig01b}
where the even-symmetry states, the ``pure barrier-billiard states'', are 
expected to show the generic behavior. 
We provide analytical and extensive numerical calculations showing that the 
spectral properties are fully consistent with SP statistics.
Moreover, our results throw some light on the nonstationarity 
observed in~\cite{Gorin01,PG01}.

The paper is organized as follows. After defining the billiard family in 
\sect~\ref{sec:barrier}, we compute analytically the form factor for small 
arguments in \sect~\ref{sec:formfactor}. Numerical results are presented in
\sect~\ref{sec:numerics}. Section~\ref{sec:conc} contains a conclusion.

\section{Barrier billiards}
\label{sec:barrier}
The family of barrier billiards consists of rectangles with sizes $l_x$,
$l_y$ and a barrier placed on the symmetry line $x=l_x/2$ as shown in 
\fig~\ref{fig:barrier}(a). The length of the barrier $l \in (0,l_y)$ is the
only nontrivial parameter.    
The free motion of a point particle with mass $m$ and momentum $(p_x,p_y)$ 
bounded by elastic reflections at the boundary of the billiard has a second
constant of motion $K = p_x^2$ in addition to Hamilton's function $H$. Hence,
the dynamics in phase space is restricted to invariant surfaces $(H,K) =
\const$. The topology of these surfaces is not that of a torus (with genus 1)
but that 
of a two-handled sphere (genus 2) due to the critical corner at the upper end 
of the barrier; see~\cite{RichensBerry81} for the relation between critical
corners and the genus of invariant surfaces in pseudointegrable billiards.

The billiards are Veech (roughly speaking, this property implies a special
kind of hidden symmetry) if and only if $l/l_y$ is a rational 
number~\cite{EMS01}.  
Still, a typical symmetric barrier billiard is not a generic pseudointegrable
system since it is composed of two copies of an integrable sub-billiard, the 
rectangle shown in \fig~\ref{fig:barrier}(b). 
This property is identical to almost-integrability~\cite{Gutkin86} in the case
of $l/l_y$ being rational. 
\def\figbarrier{%
(a) Trajectory (dotted) in the full barrier billiard, a rectangle with a
barrier between the points $(x,y) = (l_x/2,0)$ and $(l_x/2,l)$. 
Symmetry-reduced system with (b) Dirichlet boundary conditions and 
(c) mixed boundary conditions:
Dirichlet (Neumann) on solid (dashed) lines.}
\def\FIGbarrier{\centerline{\includegraphics[width=8.0cm,angle=0]{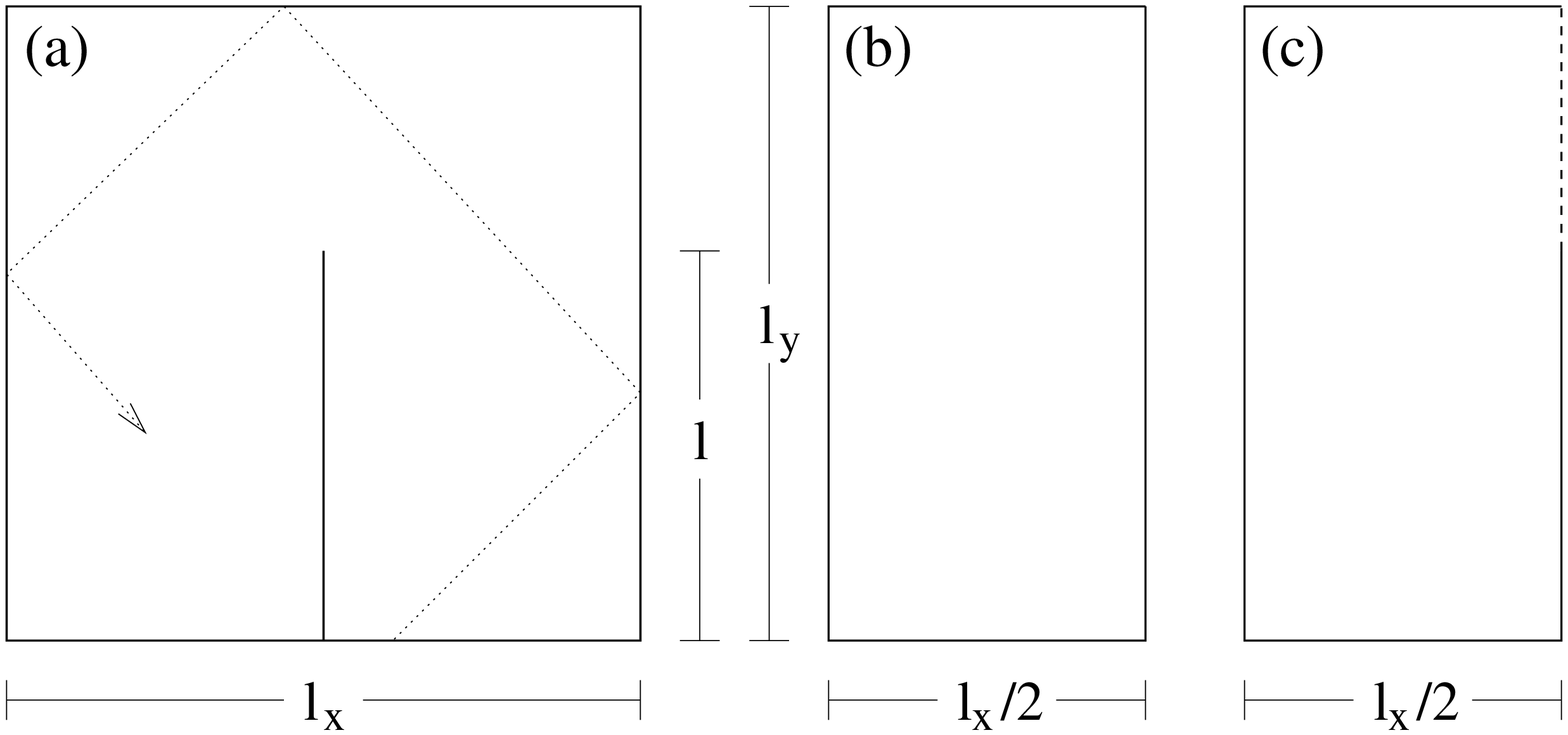}
}}
\FIGo{fig:barrier}{\figbarrier}{\FIGbarrier}

The energy eigenstates are solutions of the Helm\-holtz equation with
Dirichlet boundary conditions, i.e. vanishing amplitude, on the boundary of 
the polygon. The states are odd or even with respect to the symmetry line. 
The former ones are trivial eigenstates of the integrable sub-billiard in
\fig~\ref{fig:barrier}(b). We therefore deal mainly with the even ones, the
``pure barrier-billiard states'', which fulfill mixed boundary conditions on
the boundary of the symmetry-reduced polygon; see \fig~\ref{fig:barrier}(c).
We expect that the pure barrier-billiard states show the generic features of
energy states in rational polygons. 

\section{The spectral form factor}
\label{sec:formfactor}
We here compute analytically a spectral quantity, the 2-point correlation 
form factor, for the energy levels $\{E_n\}$ of a special member of the
barrier-billiard family. Our calculation is 
inspired by that for the triangular billiards in~\cite{BGS01}. It turns out
that the present calculation is much simpler. As in~\cite{BGS01}, we will 
apply the modern semiclassical theory based on trace formulas which 
express the density of states of a quantum system in terms of
periodic orbits of the underlying classical system~\cite{Gutz90}. 
For billiards, the semiclassical limit $\hbar\to 0$
corresponds to the high-energy limit $E\to\infty$. 
Throughout the paper we use natural units such that $\hbar = 2m = 1$. 

The density of states can be written as sum of a smooth part and an
oscillatory part 
\begin{equation}
d(E) = \sum_n\delta(E-E_n) = \bar{d}(E)+\dosc.
\end{equation}
The fluctuations in the oscillatory part can be studied with the help of the 
2-point correlation function  
\begin{equation}
R(\epsilon) = \left< {d_{\text{\footnotesize
osc}}(E+\frac{\epsilon}2)}{d_{\text{\footnotesize
osc}}(E-\frac{\epsilon}2)}  \right> \ . 
\end{equation}
Brackets $\langle\ldots\rangle$ denote an energy averaging around $E$ on an
energy window much larger than the mean level spacing $1/\bar{d}(E)$, and much
smaller than $E$. 
The Fourier transform of $R(\epsilon)$ is the spectral form
factor
\begin{equation}\label{eq:K}
K(\tau) = \int_{-\infty}^\infty \frac{d\epsilon}{\bar{d}} R(\epsilon) e^{2\pi
i \bar{d}\epsilon\tau} \ .
\end{equation}
We will concentrate on the limit $\tau\to 0$; $K(0) = 1$ for
Poisson~\cite{Berry85b}, $1/2$ for SP~\cite{BGS01} and $0$ for
GOE~\cite{Berry85b,Mehta67}.   

For two-dimensional rational polygons, the smooth part of the density of
states is semiclassically described by Weyl's law $\bar{d}=A/(4\pi)$ where $A$
is the area of the polygon; 
the oscillating part splits into two parts~\cite{VWR94,PS95}
\begin{equation}
\dosc = \dpo+\ddo \ .
\end{equation}
The periodic orbit contribution 
\begin{equation}\label{eq:dpo}
\dpo = \sum_{\text{\footnotesize p.o.}} \frac{A_p}{4\pi}
\frac{1}{\sqrt{2\pi k l_p}}e^{ikl_p-i\nu_p\pi/2-i\pi/4}+c.c.
\end{equation}
is a summation over classical (primitive and non-primitive) periodic orbits.
These orbits are marginally stable and appear always in 
one-parameter families reflecting the foliation of phase space by
two-dimensional invariant surfaces. 
$p$ labels these families; $A_p$ denotes the surface in configuration space
covered by a given family (without repetitions of primitive periodic orbits); 
$l_p$ is the (non-primitive) length of periodic orbits; the
Maslov index $\nu_p$ is here twice the number of reflections at Dirichlet
boundaries (Neumann boundaries do not contribute); $k=\sqrt{E}$ is the wave
number.

The diffractive orbit contribution $\ddo$ is a summation over orbits
starting and ending at critical corners of the polygon. 
This summation is more involved than the periodic
orbit contribution~\cite{BPS00}. In the limit $\tau\to 0$, however, the form
factor $K(\tau)$ does not depend on diffractive orbits~\cite{BGS01}. With 
this insight a formula for $K(0)$ has been derived in~\cite{BGS01}
by inserting the periodic orbit contribution~(\ref{eq:dpo}) into
\equ~(\ref{eq:K}) and employing the diagonal approximation (which is expected
to be valid for small $\tau$) yielding
\begin{equation}\label{eq:K0}
K(0) = \lim_{\tau\to 0} \frac{1}{8\pi^2\bar{d}}
\sum_{\text{\footnotesize p.o.}}\frac{|A_p|^2}{l_p}g_p^2\delta(l_p-4\pi k
\bar{d}\tau) \ , 
\end{equation}
where $g_p$ is the multiplicity of a given periodic-orbit family, i.e. the 
number of families with exactly the same lengths, and the summation is
performed over families with different lengths.

For later considerations it is helpful to repeat the evaluation of
\equ~(\ref{eq:K0}) for the simplest case, the rectangular 
billiard, as done in~\cite{BGS01}. A family of periodic orbits in a rectangle 
with sizes $a$, $b$ ($a$ and $b$ are irrationally related) and area $A=ab$
can be specified by two non-negative  
integers $m_p$ and $n_p$, denoting the number of traversals across the
billiard in the $x$- and $y$-direction, respectively. The length of each
orbit is 
\begin{equation}\label{eq:ellipse}
l_p = \sqrt{(2m_pa)^2+(2n_pb)^2} \ .
\end{equation}
The number of periodic orbits $N(l)$ up to length $l$ is the number of lattice
points in the positive $(m_p,n_p)$-quadrant inside the
ellipse~(\ref{eq:ellipse}) asymptotically giving by
\begin{equation}
N(l) = \frac{\pi l^2}{16A} \ .
\end{equation}
Due to the fact that all families cover the same area $A_p = 2A$ and have
typically the same multiplicity $g_p = 2$ (time-reversal symmetry) 
the sum~(\ref{eq:K0}) can be replaced by the following simple integral
\begin{equation}\label{eq:K0final}
K(0) = \lim_{\tau\to 0} \frac{2A^2}{\pi^2\bar{d}}
\int_0^\infty \frac{1}{l}\delta(l-4\pi k \bar{d}\tau)\frac{dN(l)}{dl} dl 
\end{equation}
which gives $K(0) = 1$ as expected for generic integrable
systems~\cite{Berry85b}. 

We now extend the previous calculation to the barrier billiard. To keep the 
calculation elementary, we restrict ourself to the special Veech case 
$l=l_y/2$. 
The odd states are eigenstates in the rectangle with width $a=l_x/2$, 
height $b=l_y$, and with Dirichlet boundary conditions, see
\fig~\ref{fig:barrier}(b), so we get $K(0)=1$ as demonstrated above.
The even states, the ``pure barrier-billiard states'', fulfill mixed boundary
conditions as shown in \fig~\ref{fig:barrier}(c). 
In the semiclassical trace formula~(\ref{eq:dpo}), the inhomogeneous boundary
conditions only influence the Maslov indices of the periodic orbits: 
a reflection at a Dirichlet boundary increases the index by two 
in contrast to a reflection at a Neumann boundary which does not change
the index. The resulting phase difference of $\pi$ between trajectories has
an analog in billiards with a magnetic flux line~\cite{BGS01} where
trajectories encircling a flux of 1/2 (in natural units) once pickup a phase
$\pi$.  

First, let us consider periodic orbits with fixed $m_p,n_p \geq 0$ and $m_p$
odd. We write $m_p = m_N+m_D$ where $m_N, m_D\geq 0$ count the number of
reflections at 
$x=l_x/2$ with Neumann or Dirichlet boundary condition, respectively. Two
cases have to be distinguished: $m_N$ even and $m_D$ odd; $m_N$ odd and $m_D$
even. The corresponding two types of orbits are related by a symmetry
transformation (ignoring the boundary conditions), the reflection at the 
line $y=l_y/2$. Hence, both types have the same $l_p$, $g_p$ and $A_p$. 
However, the Maslov indices are different due to the
inhomogeneous boundary conditions: $\nu_p\mod = 0$ and $\nu_p\mod =
2$, respectively. This implies that the contribution of both families to the
trace formula~(\ref{eq:dpo}) are identical differing just by a sign. Therefore
both contributions cancel each other.  

Second, let us turn to periodic orbits with $m_p$ even. We begin with
unfolding the orbits into a larger rectangle with width $a' = 2a$. Assume, for
simplicity, that $m_p' = m_p/2$ is odd. The case $m_p'$ even can be treated by
further unfolding of the orbits.
Again, for $m_p' = m_N'+m_D'$ odd there exists two kinds of periodic orbits
related by symmetry: one with $m_N'$ even and $m_D'$ odd and one with $m_N'$
odd and $m_D'$ even.  
These two kinds of trajectories either become congruent or remain separated
when folded back into the original rectangle with width $a$. In the first
case, we have to add the two different values of $m_N'$ and the two values of
$m_D'$ leading to $m_N$ and $m_D$ odd throughout the family. 
In the other case, we get $m_N = 2m_N'$ and $m_D=2m_D'$ even since each orbit
is symmetric with respect to the folding axis.  
Clearly, for fixed $n_p$ only one of these two cases is possible. Hence, no
cancellation occurs for even $m_p$ in the trace formula~(\ref{eq:dpo}), in contrast to
the complete cancellation in the case of $m_p$ odd.   
The simple consequence of which is that the number of periodic-orbit families
which contributes to the trace formula~(\ref{eq:dpo}) is
reduced by a factor two. The same is true for the sum~(\ref{eq:K0}). 
From \equ~(\ref{eq:K0final}) follows then directly our main analytical result
\begin{equation}\label{eq:K0result}
K(0)=\frac12 \ .
\end{equation}
Our calculated $K(0)$ is not only close to the SP prediction as in the case of
 Veech triangles~\cite{BGS01}, it agrees exactly with the SP prediction.   

The calculation for general barrier length is considerably more
complicated. Yet, it should be possible to compute $K(0)$ also for rational
$l/l_y\neq 1/2$ using methods developed in~\cite{BGS01,EMS01}.

\section{Numerical results}
\label{sec:numerics}
We here present numerical results on several statistical quantities for general
symmetric barrier billiards. As representatives we choose the Veech billiard
with $l=l_y/2$ 
and one which is not Veech with $l=l_y\omega$, where $\omega=(\sqrt{5}-1)/2$ 
is (the reciprocal of) the golden mean. Irrationally related parameters $l_x =
\pi\sqrt{8\pi}/3$ and $l_y = 3\sqrt{8\pi}/\pi$ are taken. Billiards with
$l\approx 0$ and $l \approx l_y$ are not investigated since the semiclassical
behavior of these limiting cases is expected to set in at extremely high
energies.   
We consider two different energy regimes: 
(i) the medium-energy regime starting with the $40\,000$th level and ending
with the $60\,000$th level, and (ii) the high-energy regime starting with the
$400\,000$th level and ending with the $420\,000$th level. Our high-energy 
regime is below that of Ref.~\cite{PG01} and above that of
Ref.~\cite{Gorin01}.  

We compute the eigenvalues with the mode-matching technique which is very 
efficient for barrier billiards as described in detail in~\cite{Wiersig01}. 
An accuracy of about $10^{-4}$ of the mean level spacing is achieved.  

To distinguish between local fluctuations in the level sequence  $E_1 \leq E_2
\leq E_3\leq\ldots$ and a systematic global energy dependence of the average
density we ``unfold'' the spectra in the usual way by setting $\tilde{E}_n =
\bar{N}(E_n)$; see, e.g.,~\cite{Haake91}. $\bar{N}(E)$ is the smooth part of the 
integrated density of states $N(E)=\int d(E')dE'$ (number of levels up to
energy $E$).  
In contrast to our semiclassical analysis in  the previous section,
we have to take into consideration that our energy regime is finite, therefore
we approximate  
$\bar{N}(E)$ by the generalized Weyl's law including perimeter and corner 
corrections~\cite{BH76}. We obtain for the rectangle with Dirichlet boundary
conditions in \fig~\ref{fig:barrier}(b)
\begin{equation}
\bar{N}(E) = E-\frac{l_x+2l_y}{4\pi}\sqrt{E}+\frac{4}{16}
\end{equation}   
and for the rectangle with mixed boundary conditions in
\fig~\ref{fig:barrier}(c) 
\begin{equation}\label{eq:Nbarrier}
\bar{N}(E) = E-\frac{l_x+2l}{4\pi}\sqrt{E}+\frac{1}{16} \ .
\end{equation}    
By construction, the unfolded spectra $\{\tilde{E}_n\}$ have unit mean level
spacing. Henceforth, the tilde~$\tilde{}$ will be suppressed.   

\subsection{Nearest-neighbor spacing distributions}
An important statistical quantity measuring short-range level correlations is 
the nearest-neighbor spacing distribution. It is defined as the probability
density of the spacing $s$ between adjacent levels
\begin{equation}
P(s) = \lim_{n\to\infty}\frac{1}{n}\sum_{i=1}^n\delta(s-E_{i+1}+E_i) \ .
\end{equation}
We will compute its integral, the cumulative spacing distribution
\begin{equation}\label{eq:cnn}
I(s) = \int_0^s P(s')ds' \ .
\end{equation}
For Poisson statistics $P_{\text{\footnotesize P}}(s) = \exp{(-s)}$ and 
$I_{\text{\footnotesize P}}(s) = 1-\exp{(-s)}$, the GOE is well described by the Wigner
surmise $P_{\text{\footnotesize W}}(s) = (\pi/2)s\exp{(-\pi s^2/4)}$ and 
$I_{\text{\footnotesize W}}(s) = 1-\exp{(-\pi s^2/4)}$, and for the SP 
statistics~\cite{HFS99,BGS01b}   
\begin{equation}\label{eq:spnn}
P_{\text{\footnotesize SP}}(s) = 4s e^{-2s}, \,\, 
I_{\text{\footnotesize SP}}(s) = 1-(2s+1)e^{-2s} \ .
\end{equation}
$P_{\text{\footnotesize SP}}(s)$ shows a linear increase at small $s$ (level repulsion) like the
Wigner surmise and an exponential fall-off at large $s$ like Poisson
statistics.  

In \fig~\ref{fig:cnn} one sees that the cumulative spacing
distribution is in good agreement with the SP statistics in
both energy regimes (the medium-energy behavior of the
non-Veech billiard is not shown since it is  similar to the Veech case). 
However, small fluctuations around SP can be observed in the magnification. 
The fluctuations decrease with increasing energy, and they are larger 
than the statistical fluctuations $<0.5/\sqrt{W} \approx 0.0035$ due to the
finite width $W = 20\,000$ of the energy windows.   
For the Veech billiard, we find a slight tendency towards the
Wigner surmise for medium energies and a slight tendency towards the
Poisson distribution for high energies. The fluctuations in the non-Veech 
case are of the same magnitude but without clear tendency
towards Wigner surmise or Poisson distribution.  
\def\figcnn{%
Difference between the cumulative spacing distribution of the pure 
barrier-billiard levels. Below: magnification.} 
\def\FIGcnn{\includegraphics[width=7.5cm,angle=0]{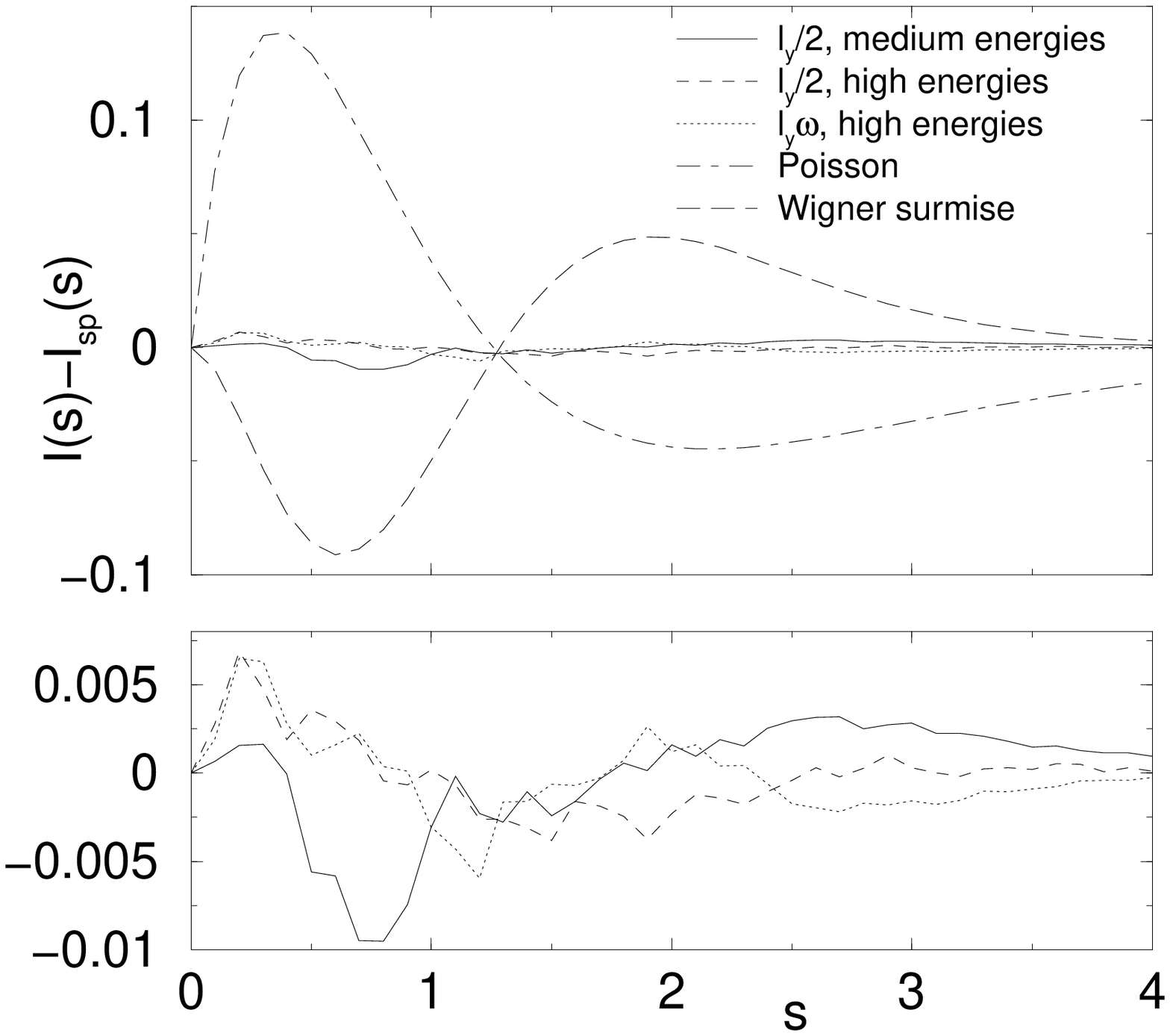}
}
\FIGo{fig:cnn}{\figcnn}{\FIGcnn}

The fluctuations for the Veech barrier billiard are very
similar (but by a factor 2.5 smaller in the high-energy regime) than those
found in right triangles~\cite{PG01}.
In~\cite{PG01}, increasing fluctuations have been reported for very high 
energies above the $4\,000\,000$th level. These
fluctuations have been interpreted as deviations from SP leading to the
conclusion that SP is asymptotically not the relevant statistics for
pseudointegrable systems.  
In the following paragraphs, however, we will show that this interpretation is
doubtful.

Let us construct an artifical SP distributed sequence of numbers. 
Take the levels of the simple rectangle in \fig~\ref{fig:barrier}(b) given by 
\begin{equation}\label{eq:E0}
E^0_{mn} = (2\pi m/l_x)^2+(\pi n/l_y)^2 
\end{equation}
with $m,n=1,2,3,\ldots$.
It has been demonstrated numerically that the nearest-neighbor spacing
distribution and some other statistical properties of such a sequence are 
asymptotically extremely well described by the Poisson
statistics~\cite{RobnikVeble98}; see also~\cite{Marklof98}.  
After ordering the levels according to increasing energy and removing every 
other level, the nearest-neighbor spacing distribution of the sequence thus
obtained obeys SP statistics~\cite{HFS99}. Figure~\ref{fig:cnnart} 
shows the corresponding cumulative spacing distribution computed numerically
from $20\,000$ levels in three different regimes. The medium- and high-energy
regime are defined as before, whereas the very-high-energy regime starts at
the $4\,000\,000$th level as in Ref.~\cite{PG01}. We observe small fluctuations
around SP which decrease with increasing energy.
In the medium- and high-energy regime, the fluctuations  
are of the same order of magnitude as for the pure barrier-billiard levels; 
cf.~\fig~\ref{fig:cnn}. We note that the same fluctuations are also
present when $I(s)-I_{\text{P}}(s)$ is plotted for the Poisson sequence given
by \equ~(\ref{eq:E0}).  

The statistical fluctuations depend on the number of levels under
consideration. This carries over to the total fluctuations as 
illustrated for the very-high-energy regime in 
\fig~\ref{fig:cnnart} with $10\,000$ and $20\,000$ levels, respectively. 
Hence, one should not compare the statistics of sequences with different
number of levels as it has been done in~\cite{PG01}.  

Following the same reasoning as described above we have also constructed a 
SP sequence of $20\,000$ numbers using a conventional 
pseudo-random number generator. 
This reproduces the expected statistical fluctuations of order $0.0035$.
To summarize, from the fluctuations found numerically here and
in~\cite{PG01,Gorin01} it is not justified to exclude SP as correct
statistics for generic pseudointegrable systems.  
\def\figcnnart{%
Difference between the cumulative spacing distribution of the artifical
SP sequence.}
\def\FIGcnnart{\includegraphics[width=7.5cm,angle=0]{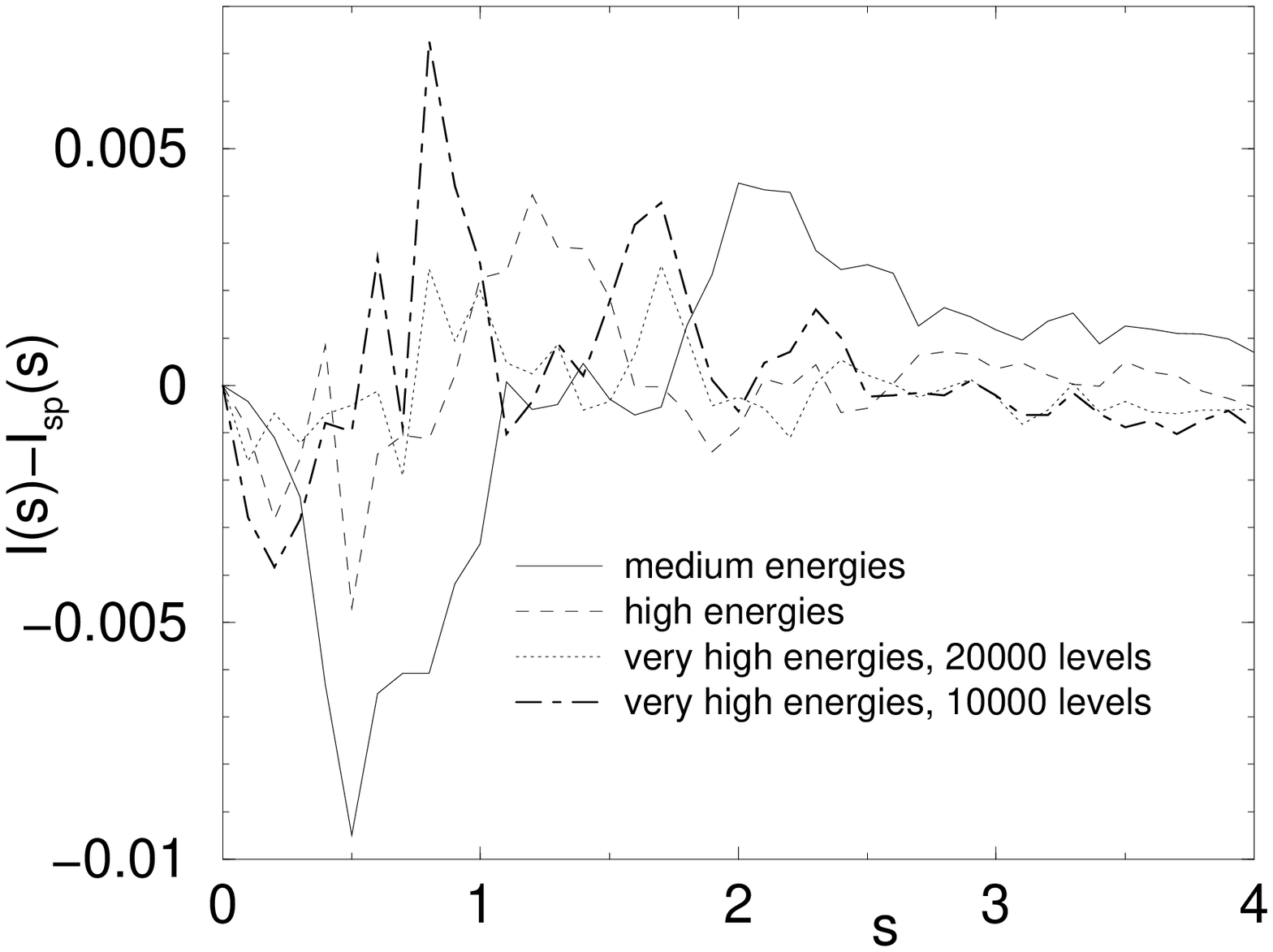}
}
\FIGo{fig:cnnart}{\figcnnart}{\FIGcnnart}

We mention that the distribution of spacings between
neighboring eigenvalues of the S-matrix in an open version of the barrier
billiard~\cite{ESTV96} also resembles the SP result; although the agreement 
is not as good as here.  

\subsection{Next-to-nearest spacing distributions}
In the previous subsection we have seen that the nearest-neighbor distributions
are close to the SP prediction in
\equ~(\ref{eq:spnn}). However, \equ~(\ref{eq:spnn}) is also valid for IP 
statistics.
In order to distinguish between IP and SP statistics one has to consider 
other correlation functions. First, we choose the next-to-nearest spacing 
distribution (second-neighbor-spacing distribution) and its integral. For the
SP statistics~\cite{BGS01b}   
\begin{eqnarray} 
P_{\text{\footnotesize SP}}(2,s) & = & \frac83s^3 e^{-2s}\,, \nonumber\\
I_{\text{\footnotesize SP}}(2,s) & = & 1-\frac13(4s^3+6s^2+6s+3)e^{-2s} \ .
\end{eqnarray}
For IP we find analytically
\begin{eqnarray} 
P_{\text{\footnotesize IP}}(2,s) & = & 4e^{-s}[1-(1+s)e^{-s}]\,, \nonumber\\
I_{\text{\footnotesize IP}}(2,s) & = & 1+e^{-s}[e^{-s}(3+2s)-4] \ .
\end{eqnarray}

Figure~\ref{fig:nextcnn} shows that the cumulative next-to-nearest spacing
distribution is in agreement with the SP statistics but not with IP
statistics.
Note that the fluctuations are two times larger as in the case of the
nearest-neighbor spacing distribution in \fig~\ref{fig:cnn}.  
\def\fignextcnn{%
Difference between the cumulative next-to-nearest spacing distribution of the
pure barrier-billiard levels.}
\def\FIGnextcnn{\includegraphics[width=7.5cm,angle=0]{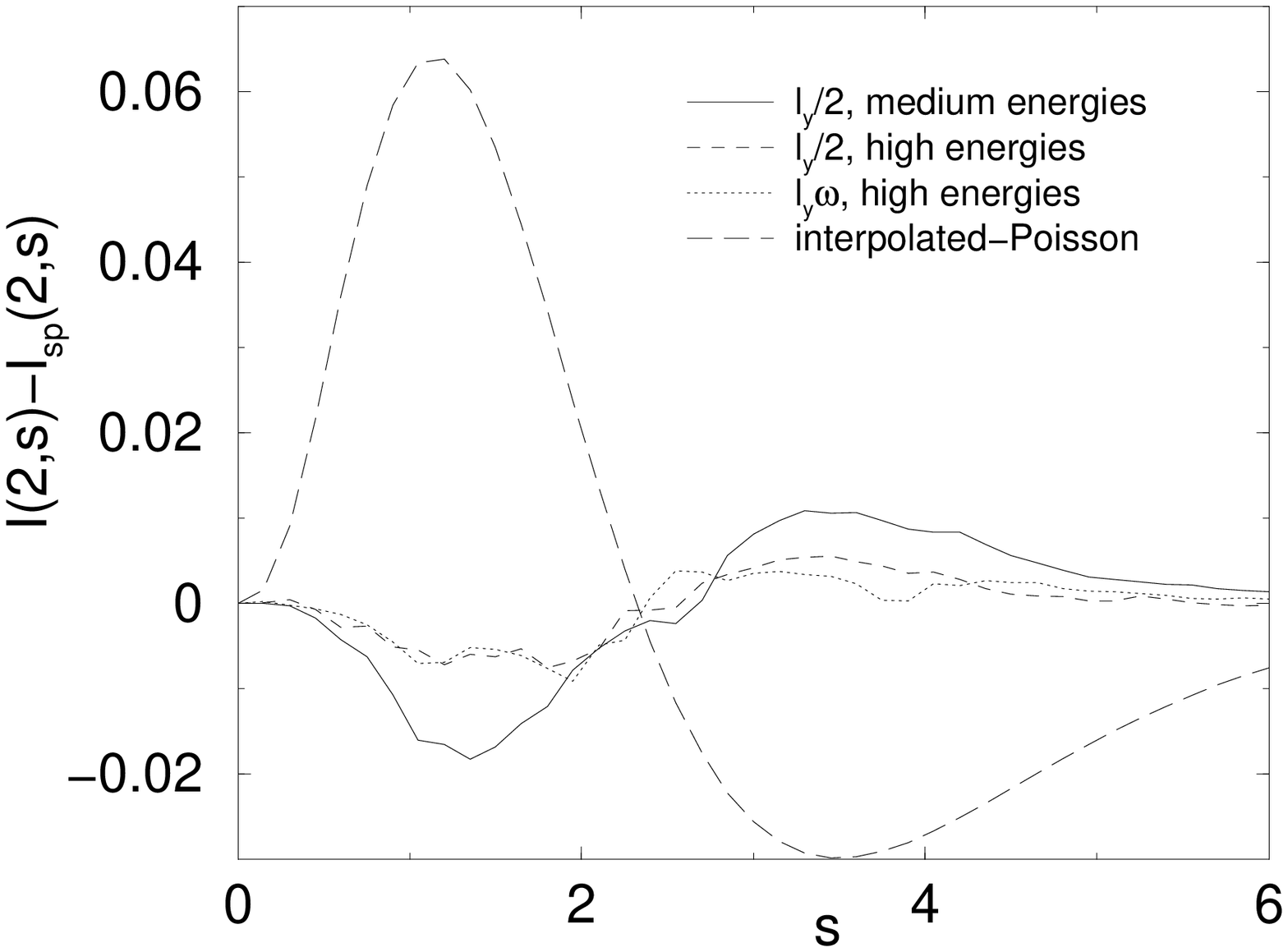}
} 
\FIGo{fig:nextcnn}{\fignextcnn}{\FIGnextcnn}

We have also investigated $n$th-neighbor spacing distributions $P(n,s)$ with
$n=3,4,5$. Again, the distributions differ significantly from IP statistics
and are well described by SP statistics, even though the fluctuations 
increase slightly. A detailed discussion is left out since we will study
long-range level correlations in a more comprehensive way in the next
subsection.   

\subsection{Number variance}
The number variance
\begin{equation}\label{eq:sigma}
\Sigma(L) = \left<(n(L,E)-L)^2\right>
\end{equation} 
is the local variance of the number $n(L,E) = N(E+L/2)-N(E-L/2)$ of
energy levels in the interval $[E-L/2,E+L/2]$. SP statistics
gives~\cite{BGS99,HFS99,BGS01b} 
\begin{equation}
\Sigma_{\text{\footnotesize SP}}(L) = 
\frac{L}{2}+\frac{1}{8}(1-e^{-4L}) \ .
\end{equation}
For IP statistics we get analytically a different result
\begin{equation}
\Sigma_{\text{\footnotesize IP}}(L) = 
L-\frac{1}{2}+(L+\frac{1}{2})e^{-2L} \ .
\end{equation}

Figure~\ref{fig:sigma} reveals a substantial difference to SP for 
correlation lengths $L > 4$ in the medium-energy regime. In the
high-energy regime the difference is smaller. 
Note that the $L$ regime in \fig~\ref{fig:sigma} is well below the crossover
region where the number variance begins to saturate at a value determined 
by the shortest periodic orbit~\cite{Berry85b}.
In the region of large $L$, the number variance is related to the
form factor (see, e.g.,~\cite{BGS01}) by means of
\begin{equation}\label{eq:asymsigma}
K(0) = \lim_{L\to\infty} \frac{\Sigma(L)}{L} \ .
\end{equation}
Using this relation we get
for the Veech case $K(0) \approx 0.27$ at medium energies and $\approx
0.34$ at high energies ($\approx 0.36$ for the non-Veech case). 
However, we do not interpret this result as deviation from SP since we know
from Section~\ref{sec:formfactor} that in the Veech case $K(0)$ does
converge to the SP result $1/2$. Hence, we conclude that the convergence to a
stationary limit is extremely slow.
The slow convergence of the spectral statistics is shared by related systems 
such as right triangular billiards~\cite{BGS01,Gorin01,PG01}, rectangular
billiards with magnetic flux lines~\cite{BGS01,RF01}, and parabolic maps with 
spin~\cite{HK01}.  
To overcome the problem of slow convergence, we have tried to use an
extrapolation procedure described in~\cite{BGS01}. However, in our case it 
does not give satisfactory results and therefore a detailed discussion is
omitted. 
\def\figsigma{%
Number variance $\Sigma(L)$ of the pure barrier-billiard levels.} 
\def\FIGsigma{\includegraphics[width=7.5cm,angle=0]{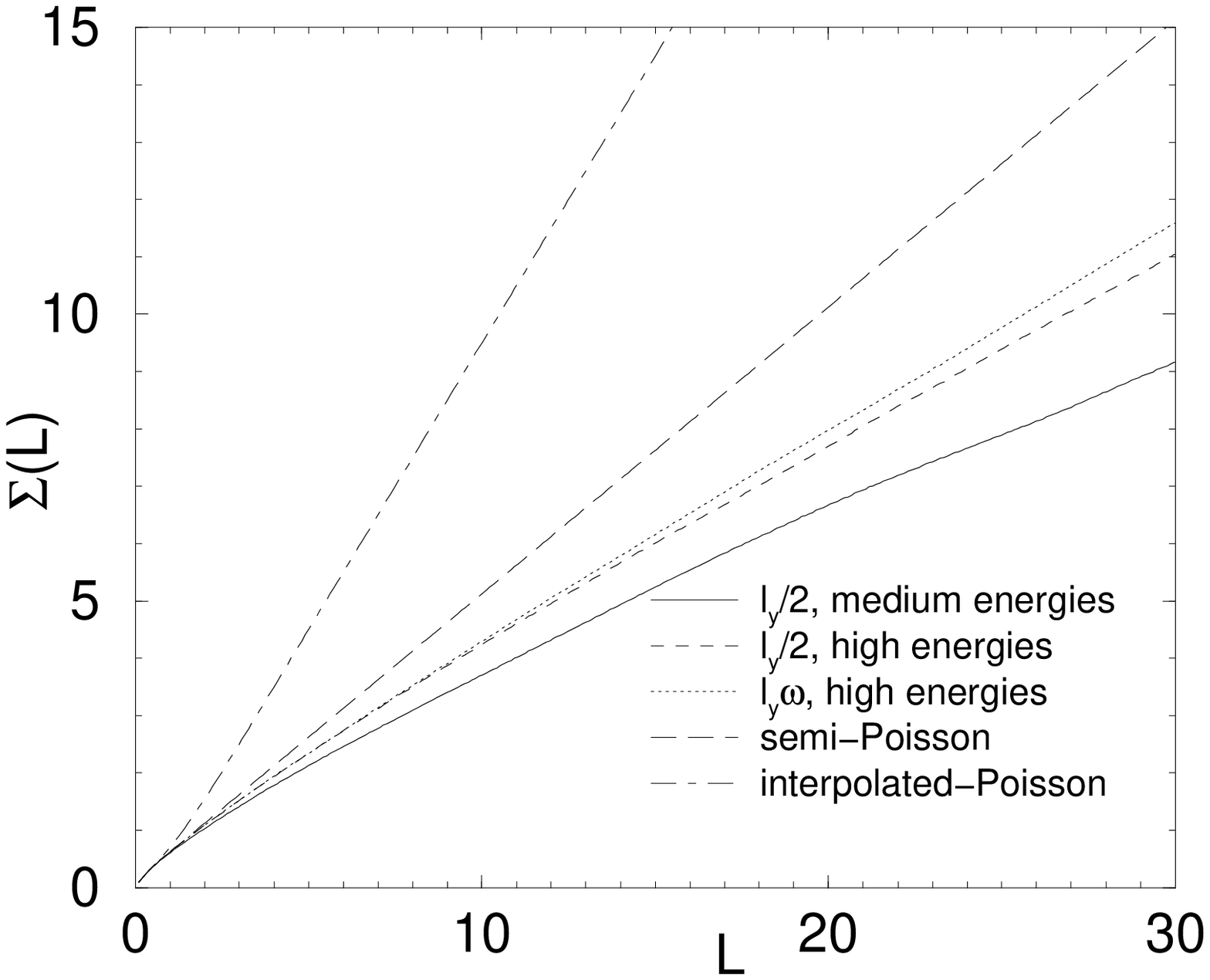}
}
\FIGo{fig:sigma}{\figsigma}{\FIGsigma}

Figure~\ref{fig:sigmaart} shows the number variance for the artifical SP
sequence 
constructed from the levels of the integrable rectangle. The convergence in
direction towards SP is similar, even though a bit faster, as for the pure
barrier-billiard levels plotted in \fig~\ref{fig:sigma}. In the regime of very
high energies, the number variance is hard to distinguish from the SP
curve.
\def\figsigmaart{%
Number variance $\Sigma(L)$ of the artifical SP sequence.} 
\def\FIGsigmaart{\includegraphics[width=7.5cm,angle=0]{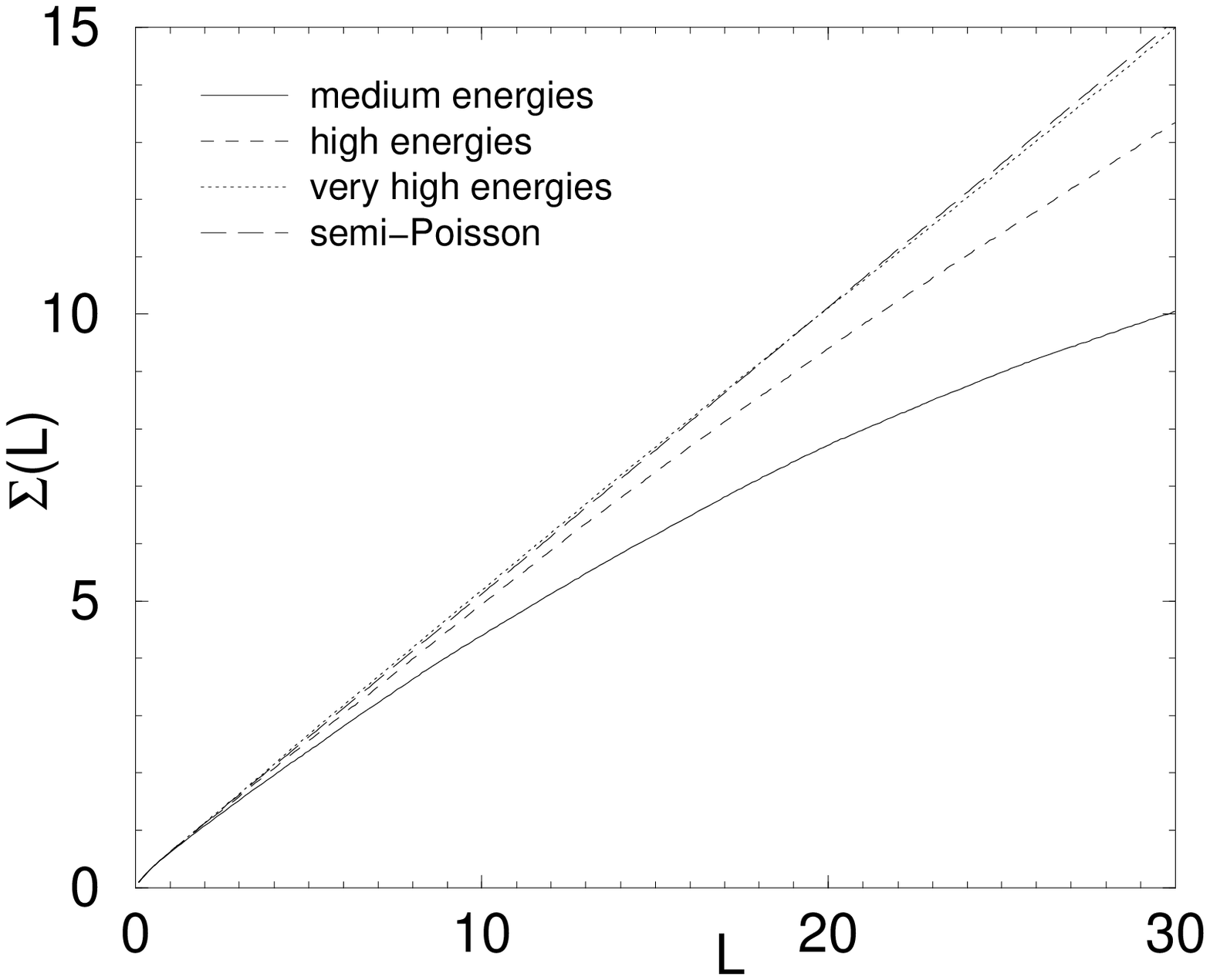}
}
\FIGo{fig:sigmaart}{\figsigmaart}{\FIGsigmaart}

\subsection{The form factor}
The form factor $K(\tau)$ can be approximated numerically by (see, e.g.,
Ref.~\cite{Marklof98})  
\begin{equation}\label{eq:Ktn}
K(\tau;n) =  \frac{1}{n}\left|\sum_{j=l}^{l+n} e^{2\pi i E_j\tau}\right|^2 .  
\end{equation}
We here consider only the high-energy regime, i.e. $l = 400\,000$ and 
$n=20\, 000$. We smooth $K(\tau;n)$ by averaging over small intervals of 
size $\Delta\tau = 0.006$. Nevertheless, the numerical data is quite
irregular as can be seen in \fig~\ref{fig:Kav} for the Veech billiard (for the
non-Veech case the picture looks very similar). It is difficult to estimate 
$K(0)$ directly from such kind of data, but it is clear that $K(0)$ is well 
below the SP prediction $1/2$, which is consistent with our former numerical 
results on the number variance. 

A more elegant way to compare the form factor to SP statistics is 
described in Ref.~\cite{BGS01}. Fit $K(\tau;n)$ to the function 
\begin{equation}\label{eq:Kfit}
K_{\text{fit}}(\tau) = \frac{a^2-2a+4\pi^2\tau^2}{a^2+4\pi^2\tau^2} \ .
\end{equation}
Expression~(\ref{eq:Kfit}) is the SP form factor when $a=4$. Therefore,
the quantity $K_{\text{fit}}(0)-1/2$ measures the difference to SP
statistics. Note that $K_{\text{fit}}(0)$ in general differs from $K(0;n)$
since it depends also on $K(\tau;n)$ with $\tau > 0$. 
Fitting \equ~(\ref{eq:Kfit}) to our smoothed data over the range
$0\leq\tau\leq 3$, we find remarkable agreement with SP statistics: 
$K_{\text{fit}}(0) = 0.504$ for the Veech billiard (see \fig~\ref{fig:Kav}) 
and $K_{\text{fit}}(0) = 0.498$ for the non-Veech billiard. 
\def\figKav{%
The form factor~(\ref{eq:Ktn}) of the pure barrier-billiard levels;
$l=l_y/2$. The smooth curve is the fit~(\ref{eq:Kfit}) with $a=4.032$.}
\def\FIGKav{\includegraphics[width=7.5cm,angle=0]{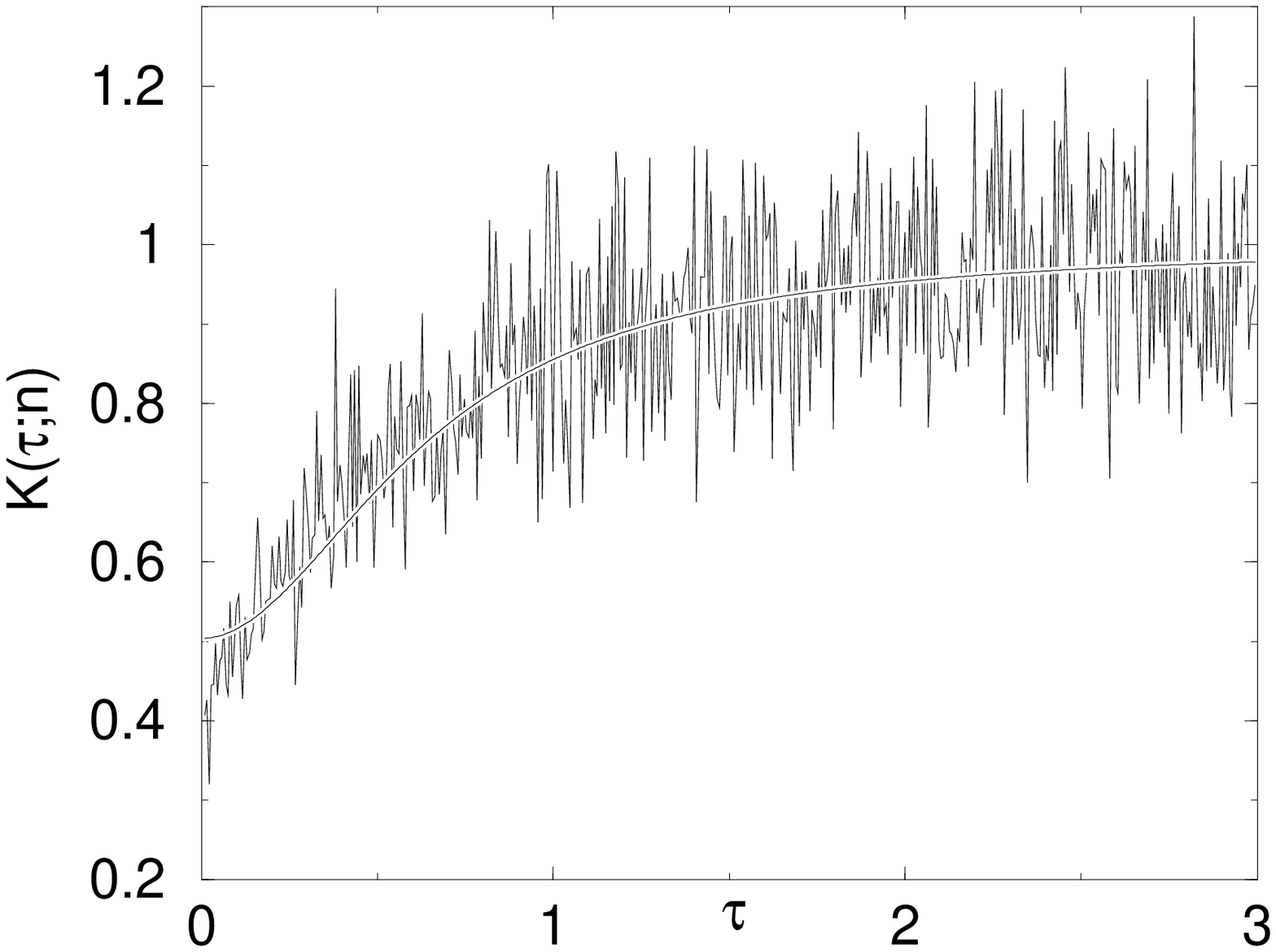}
}
\FIGo{fig:Kav}{\figKav}{\FIGKav}

\subsection{Level dynamics}
We here investigate the dependence of the energy levels on the system
parameter $l/l_y$. This so-called ``level dynamics'' has been intensively
studied for classically integrable and chaotic systems; see 
e.g.~\cite{Haake91}. To the author's knowledge, only one pseudointegrable (for
all parameter values) example, the ``square torus 
billiard''~\cite{RichensBerry81} and a generalized version of
it~\cite{SimmelEckert95}, has been studied in this regard.

A typical situation for the pure barrier-billiard levels is displayed in 
\fig~\ref{fig:dynamics}. The global increase of the levels (not
unfolded) is due to the fact that the smooth part of the integrated density of
states in \equ~(\ref{eq:Nbarrier}) decreases as $l$ is increased.
Apart from this rather trivial fact we observe a number of interesting
features: 
(i) the levels tend to avoid each other. Closer examination of the available 
numerical data indicates that there are no level crossings. That means for 
fixed parameter value there are no degeneracies in the spectrum, which is 
consistent with the SP and the GOE prediction for the nearest-neighbor spacing
distribution $P(0) = 0$ in agreement with our former numerical results.   
The total absence of level crossings is in contrast to the situation in the 
square torus billiard where crossings can appear for parameter values at which
the billiard is almost integrable~\cite{RichensBerry81}. 
(ii) Large areas free of levels exist similar as in integrable systems and 
different to fully chaotic systems. This is consistent with Poisson and SP
statistics which both predict a slower fall-off of $P(s)$ at large $s$ than 
GOE statistics does. 
(iii) There exists an unusual structure of plateaus interrupted by steep 
segments not only near avoided crossings but also fairly far away from 
avoided crossings. 
Observation of the energy eigenfunctions reveals that plateaus (steep segments)
correspond to parameter values at which the corresponding eigenfunction has 
small (large) amplitude at the upper end of the barrier. Hence, varying the
barrier length has no (strong) influence on the wave pattern and on the energy,
resulting in a plateau (steep segment).
\def\figdynamics{%
Pure barrier-billiard levels as functions of the barrier
length $0.05l_y\leq l \leq 0.95l_y$.} 
\def\FIGdynamics{\includegraphics[width=7.5cm,angle=0]{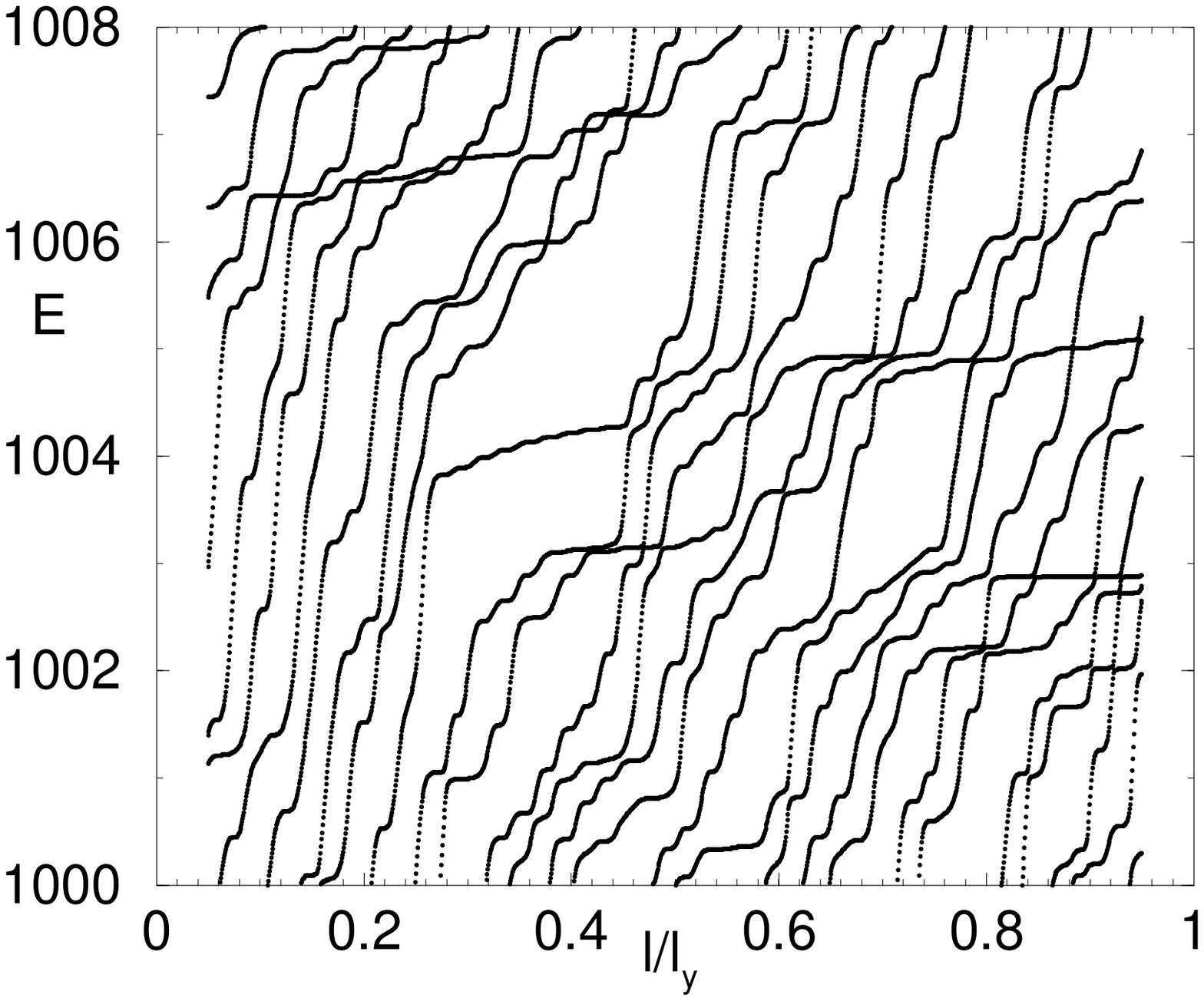}
}
\FIGo{fig:dynamics}{\figdynamics}{\FIGdynamics}

Interestingly, the abrupt changes in the slopes 
fairly far away from avoided crossings can be simulated in a natural way by 
an artifical SP sequence constructed as described before by removing every 
other level of an integrable system. Two
neighboring levels of an integrable system typically cross each other when a 
parameter is varied as sketched in \fig~\ref{fig:crossing}. 
Removing the second level (measured from below for each value of the 
parameter) gives the solid, nondifferentiable line. This could produce the
kind of abrupt changes seen in \fig~\ref{fig:dynamics}.
Of course, the slope of finite-energy levels cannot change discontinuously. 
Real discontinuouities can only be expected in the semiclassical limit. 
\def\figcrossing{%
Sketch of the local level dynamics of the artifical SP sequence.} 
\def\FIGcrossing{\includegraphics[width=5.5cm,angle=0]{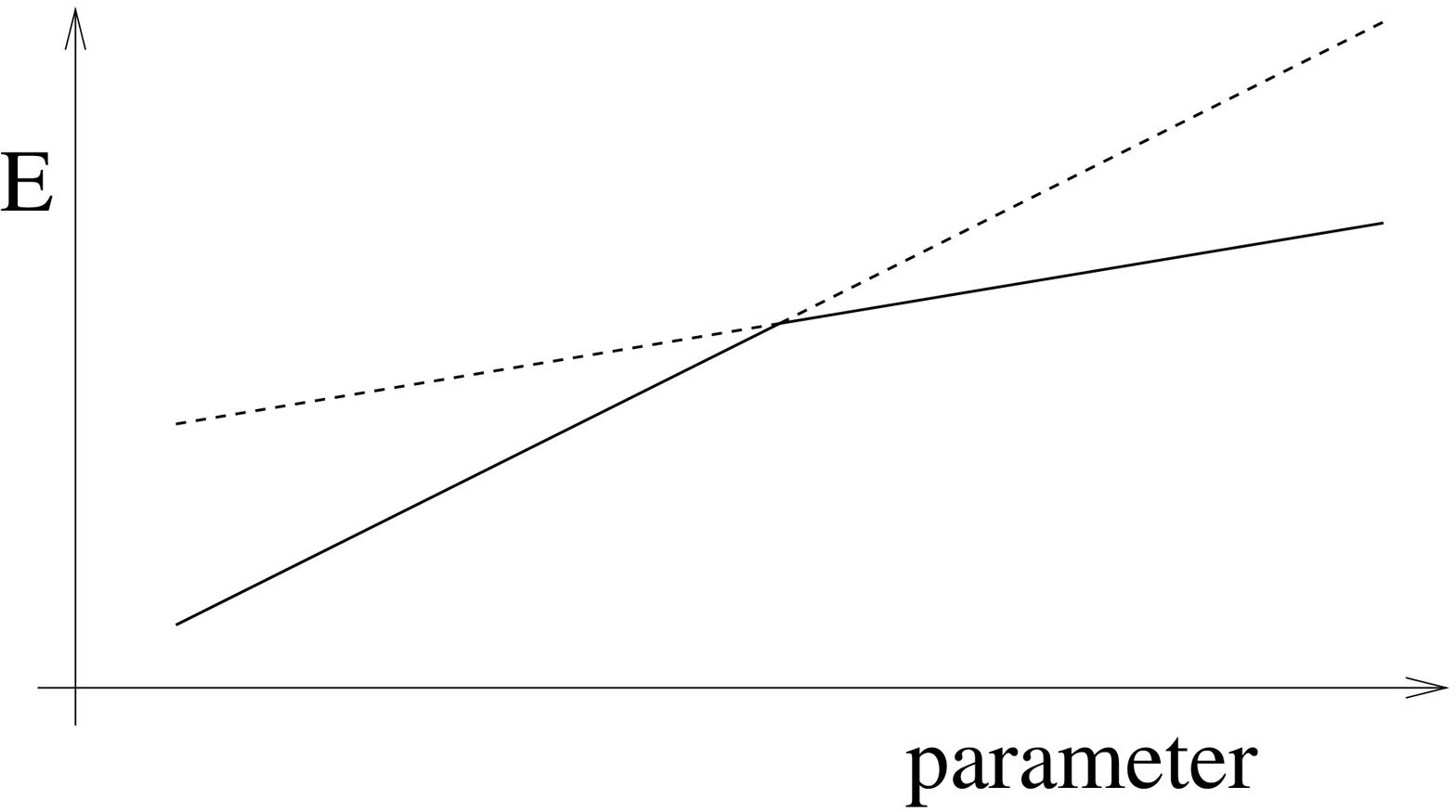}
}
\FIGo{fig:crossing}{\figcrossing}{\FIGcrossing}

\section{Conclusion}
\label{sec:conc}
In this paper we have studied the energy levels of pseudointegrable
barrier billiards. Focusing on the pure barrier-billiard states, 
we have found numerically that the nearest-neighbor spacing distributions and
next-to-nearest spacing distributions agree with the semi-Poisson (SP) 
statistics which is obtained by dropping every other number from a random
sequence. The number variance and the spectral form factor agree with SP, even
though long-range correlations seem to converge rather slowly. Moreover, the 
level dynamics is consistent with SP statistics. 
Even though we have considered an high-energy window 
($20\,000$ levels starting at the $400\,000$th level) we cannot exclude that at
larger energies a different scenario takes place. However, our analytical
result for the spectral form factor for a Veech barrier billiard,
$K(\tau)\to 1/2$ as $\tau\to 0$, 
gives us some confidence that the spectral statistics of barrier billiards
are indeed close to SP.   

Due to the slow convergence of the spectral statistics in polygonal billiards,
and other diffractive systems, semiclassical methods as shown here and
in~\cite{BGS01} have to be extended in the future to higher order in $\tau$
(as in~\cite{BG01} for rectangular billiards with point-like singularities)
and to other polygons in order to clarify the role of the SP statistics in 
pseudointegrable systems.  
\rem{Of special interest is the relation to global classical properties, such
as the genus of the invariant surfaces and the ergodic properties on these 
surfaces.}

\begin{acknowledgments}
I would like to thank T. Gorin, A. B\"acker, S. Keppeler, H. Schomerus,
E. Bogomolny, C. Schmit, S. Fishman, and M. Schmoll for
discussions and comments, G. Casati and T. Prosen for communicating
their results before publication. 
\end{acknowledgments}

\bibliographystyle{prsty}
\bibliography{}

\end{document}